# Orbital ordering, ferroelasticity, and the large pressure induced volume collapse in PbCrO$_3$

P. Ganesh[1] and R. E. Cohen[1]

[1] Geophysical Laboratory, Carnegie Institution of Washington, Washington, DC 20015, USA

**Abstract:**

We report a new tetragonal ground-state for perovskite-structured PbCrO$_3$ from DFT+U calculations, and explain its anomalously large volume. The new structure is stabilized due to orbital ordering of Cr-d in the presence of a large tetragonal crystal field, mainly due to off-centering of the Pb atom. At higher pressures (smaller volumes) there is a first-order transition to a cubic phase where the Cr-d orbitals are orbitally liquid. This phase-transition is accompanied by a ~11.5% volume collapse, one of the largest known for transition-metal oxides. The large ferroelasticity and its strong coupling to the orbital degrees of freedom could be exploited to form potentially useful magnetostrictive materials.

Transition metal oxides show complex and interesting behavior due to the electrons in the *d* orbitals. Depending on the crystal structure and their magnetic state, they could be metallic or insulating or undergo a transition from one to the other[1]. The *d*-electrons are known to show itinerant as well as a strongly localized behavior. Correlated band-theory methods with the Hubbard U parameter[2,3] to capture strong correlations in the Cr d-orbitals in CrO$_2$[4,5], where Cr has a large nominal valence of 4+ with 2 electrons in the d-manifold, show that it is a half-metallic ferromagnet, with the metallicity coming from a negative charge transfer i.e. holes in the oxygen band. Chromites in the perovskite structure such as PbCrO$_3$, CaCrO$_3$ and SrCrO$_3$ have been little studied because high-pressures and temperatures are required to synthesize[6] them. CaCrO$_3$, an antiferromagnet, was long considered to be an insulator, but recent infrared reflectivity and transport measurements[7,8] show that it is a metal. For over 40 yrs. SrCrO$_3$ was thought to be a paramagnetic metallic oxide with a cubic perovskite structure, but recent neutron and x-ray diffraction experiments identified weak strain induced magnetism[9,10] which was recently shown to be stabilized by orbital ordering[11]. Further resistivity measurements suggest a pressure induced insulator-metal transition in SrCrO$_3$[10]. At room temperature PbCrO$_3$ is reported to have a semiconducting band-gap[6]. Density functional studies of PbCrO$_3$ are few[12,13], but suggest presence of strong correlations in the Cr *d*-manifold and find it to be metallic.

Whereas PbVO$_3$[14] and PbTiO$_3$[15] are ferroelectric insulators with a large tetragonal strain, x-ray and neutron diffraction studies of PbCrO$_3$[6,16-18] show that it has a cubic structure. The ionic radii[19] in an octahedral environment for Ti, V, Cr and Mn are 0.605, 0.580, 0.550 and 0.530 Å, respectively. The pseudocubic lattice parameters of their corresponding Pb compounds in the perovskite structure are 3.980 Å, 4.075 Å, 4.010 Å and 3.877 Å (high pressure) respectively. In spite of a small ionic radius of Cr, $PbCrO_3$ shows an increased lattice parameter compared to $PbTiO_3$ and is almost comparable to $PbVO_3$. Even among the chromites, $PbCrO_3$ has the largest lattice parameter, larger than $SrCrO_3$ (a~3.82 Å) even though the ionic radius of Sr (1.32 Å) is comparable to that of Pb (1.33 Å). The lattice parameter is more comparable to that of $BiCrO_3$ (a~3.9 Å), which is large due to a highly distorted polar-space group with multiferroic properties. The increased lattice parameter of $PbCrO_3$ is thus quite puzzling and indicates presence of a nearby structural phase-transition. Indeed, recent high-resolution transmission electron microscopy suggests existence of Pb displacements[18] up to ~0.29Å in $PbCrO_3$.

We performed structural optimization, relaxing atomic positions and strain, using DFT+U[2, 3] methods in the fully-localized limit[3, 20, 21] to compute the ground-state of ordered $PbCrO_3$. The DFT+U method describes the on-site Coulomb interaction of the strongly correlated d-electrons with a Hubbard like term. This method is known to give band-gaps and accurate magnetic moments in transition metal oxides including $CrO_2$[4] where the local moment is ~ 1.9 $\mu_B$, similar to that in $PbCrO_3$[16]. We chose a value of U=3eV and J=0.87 eV obtained for $CrO_2$[4] where Cr has a similar valence. We also compute the equation of state for the ground state structure and find a pressure induced phase transition accompanied by a large volume collapse.

All calculations were performed with ABINIT[22, 23] using the PAW method with both LDA and PBE-GGA exchange correlation functionals. The PAW radii were 2.309 a.u for Pb, 2.303 a.u. for Cr, and 1.004 a.u. for O. A plane-wave cutoff of 40 Ha and 80 Ha for the fine FFT mesh was used with a 6x6x6 k-point mesh for cubic symmetry, and a 4x4x4 mesh for lower symmetry structures and a Gaussian smearing of 0.005 Ha was used to accelerate electronic convergence. In structural optimizations, forces were converged until the maximum force was $5 \times 10^{-6}$ Ha/Bohr and the total energy had converged to $10^{-10}$ Ha. Once the forces were converged, the k-points were increased to get the total energies. These energies for several different volumes were then fitted to a 3$^{rd}$ order Birch-Murnaghan equation to obtain the equation of state (EOS) parameters.

At its equilibrium volume in the cubic (*Pm-3m*) phase (Table-I), we find the experimentally observed G-type anti-ferromagnetic state to be lower in energy than a ferromagnetic or a non-magnetic state. The energy difference between Ferromagnetic and G-type AFM is small, ~28meV/unit cell. The local magnetic moment in the G-AFM arrangement is 2.17 $\mu_B$, close to the localized limit of 2 expected for Cr 4+ with localized d-electrons and consistent with experimental values. As such we perform all structural optimization studies for a G-type anti-ferromagnetic structure. Further, the calculated lattice parameter is much smaller than the experimental one.

To identify effects of Pb displacements at the experimental volume, we displace Pb along the (001), (110) and the (111) pseudo-cubic directions with all other atoms in their ideal cubic positions.   Figure 1 shows the total-energy as a function of Pb displacements in reduced coordinates along these directions with LDA+U.   Pb displacement along any direction lowers the energy relative to the cubic phase (i.e. zero Pb displacement).   A displacement along the tetragonal (001) direction is most preferred.   A similar plot at a 20% reduced volume (i.e. $V/V_0 = 0.8$) shows preference for an undisplaced cubic structure.   This suggests that cubic $PbCrO_3$ has polar instabilities favoring a low symmetry polar ferroelastic structure at ambient conditions.   A more centrosymmetric phase is favored at smaller volume i.e. higher pressures, indicating a pressure induced phase transition.

Pb off-centering is common in ferroelectric systems such as $PbTiO_3$ and occurs due to hybridization between Pb and O atoms and gives rise to a ferroelectric ground-state[15],but is unknown hitherto in $PbCrO_3$.   To obtain the ground state structure and identify the origin of the driving mechanism behind the polar instability we perform full optimization of the atoms in the tetragonal phase.

Figure 2 shows the equation of state plot for the cubic phase (with LDA+U and GGA+U) as well as the fully relaxed tetragonal phase (GGA+U).   The tetragonal phase is the ground-state at ambient conditions and has the same space group as that of $PbTiO_3$ i.e. *P4mm*.   The tetragonal strain is compressive (i.e. $c/a \sim 0.88$) with a strain of ~ 12% at P=0. The strain is large, and an LDA+U calculation resulted in a similar amount of strain of ~ 9% at the LDA P=0.   The high pressure phase is cubic. DFT calculations within the local spin density approximation and the PBE GGA exchange-correlation functional with U=0 and J=0 also predict a low energy tetragonal phase with ~ 15% strain at the experimental volume.   Including the effects of spin-orbit interaction gave a metallic phase in both the cubic and the tetragonal phases, with a small change in their total energy difference (~0.01 eV/unit cell) at the experimental volume

Table I lists the equation of state parameters for the two phases with GGA+U. The cubic phase has a much smaller volume than the experimental volume.   With LDA+U, the equilibrium volume of the cubic phase ($V_0 = 53.31$ Å$^3$) is lower than with GGA+U and the bulk modulus is larger ($K_0 = 210$ GPa).   The bulk moduli of the two phases are very different. The low pressure tetragonal phase has a very high compressibility compared to the high pressure cubic phase.   This large discrepancy is expected to give rise to a large volume change at the transition.   Indeed we find a 11.5% volume collapse associated with the transition (Figure 2) from our GGA+U calculations. LDA+U gives a volume collapse of ~ 8.6%.   The predicted GGA+U transition pressure is P ~ 7GPa.   The calculated volume collapse is in good agreement with recent experiments[24] which suggest a ~9.8% volume collapse with pressure.   The magnetic moment for the tetragonal ground state is ~ $2.35\mu_B$,  close to that found in the cubic phase at the experimental volume and within the experimental range of measurements[6, 16].

Reanalyzing the data presented in figure 4 of Ref. [11], (Figure 2. Inset **a**) we find that a similar volume collapse of at least ~9% (i.e. 55 Å$^3$ to 50 Å$^3$) should be expected even in SrCrO$_3$. The transition pressure is not easily identifiable from published results. More careful study of this system is warranted.

Unlike the equilibrium lattice parameter of the tetragonal phase, the equilibrium lattice parameter of the high-pressure cubic PbCrO$_3$ perovskite phase of ~3.86 Å agrees rather well with the empirical relationship of Ubic[25] which based on the ionic radii of the constituent atoms predicts a cubic lattice parameter of ~ 3.85 Å. *So what is stabilizing the large equilibrium volume in the tetragonal phase?*
In a cubic phase due to the cubic crystal-field, the d-orbitals are split into two degenerate e$_g$ levels and a three-fold degenerate t$_{2g}$ level. In the tetragonal phase, due to the tetragonal crystal-field, the t$_{xy}$ level is split to lower energies from the t$_{xz}$ and t$_{zy}$ levels of the t$_{2g}$ manifold reducing the degeneracy. Figure 3 shows the electronic band-structure plot for the majority spins in the cubic and the tetragonal phases. Fat-bands are plotted for the different Cr-d orbitals. As evidenced, in the cubic phase, the center of all the t$_{2g}$ bands appear very close in energy and are lower than that of the e$_g$ bands which lie above the Fermi level. As such the Cr d-bands are orbitally liquid in the cubic crystal-field splitting. In the tetragonal phase, the t$_{xy}$ band is pushed well below the other t$_{2g}$ orbitals, and forms an almost dispersion-less band. As such energy is reduced by fully occupying this lower band. This leads to orbital ordering, i.e. an orbitally crystalline state with G-type ordering of the t$_{xy}$ Cr-d orbital.

The large tetragonal crystal-field splitting appears to be mainly coming from Pb off-centering (see Figure 1). The crystal-field in the centrosymmetric case (i.e. *P4/mmm* phase) is very weak at V$_0$ (i.e. the occupation of the three t$_{2g}$ levels are nearly the same). But when Pb is allowed to off-center this splitting increases and the occupation of the split t$_{xy}$ level becomes close to one for the majority spin-state. As mentioned before, Pb off centering is seen in other materials such as PbTiO$_3$ due to Pb-O hybridization and is also seen experimentally in PbCrO$_3$[18]. Changing Pb positions changes the local-crystal field on chromium. But such a large effect on the crystal-field splitting as seen here is new.

The predicted volume collapse has recently been reported experimentally[26]. The observed volume collapse is 9.8% and quite large as predicted. The experimental transition pressure is reported to be ~1.6 GPa, lower than our predicted value of ~7 GPa using GG+U with a U=3 eV. But they wrongly assign the ground-state structure to be cubic. In fact from their figure 2, it is quite clear that the cubic (100) peak in their Phase-I is split into two peaks as expected for a tetragonal phase where c/a is less than one. Only after the transition at their experimental 1.6 GPa, the peak-splitting vanishes as expected in a cubic phase. In spite of the differences between theory and experiments, our calculations not only predict the experimentally observed large volume-collapse but also explain its electronic origin.

Strong orbital ordering leads to anisotropy thereby increasing electron localization effects[27, 28]. As such the origin of magnetism under ambient conditions is partly due to these strongly localized electrons. The splitting amongst the $t_{2g}$ orbitals decreases with pressure leading to a high-pressure cubic phase which is orbitally liquid, thereby reducing the effect of localization on magnetism.

The electronic structure of the two phases is metallic in all our calculations. Even with values of U as large as 10 eV the cubic phase was metallic. A 40 year old resistivity experiment[6] gives a semiconducting value of ~ $2.6 \times 10^3$ Ω cm for the resistivity at room temperature. Below 100K the resistivity shows a metallic trend, suggesting a metal insulator transition at ~100K. There is a long standing controversy regarding similar experiments on $SrCrO_3$ and $CaCrO_3$, and newer experiments[8, 9] are at odds with older measurements. Given the importance of dynamical correlations in $CrO_2$[5], they may be important in $PbCrO_3$ and open up a small semiconducting band gap with temperature. Theoretical studies based on the dynamical mean-field theory[29] as well as new spectroscopic and/or resistivity measurements are required to elucidate if $PbCrO_3$ is a metal, insulator or a semi-conductor.

We conclude that the zero pressure ground state of $PbCrO_3$ is tetragonal (*P4mm*) with large Pb off-centering. Pb-off-centering along the tetragonal direction leads to a crystal-field splitting of the Cr-d orbitals due to its reduced site symmetry, which in turn gives rise to a tetragonal ferroelastic distortion stabilizing the anomalously large volume (pseudo-cubic lattice parameter ~ 4.00 Å). As pressure increases Pb-off-centering is less favorable and as such the tetragonal-crystal field splitting is reduced. At ~7GPa there is a phase transition to a cubic phase with space group *Pm-3m* accompanied by a large volume collapse ~ 11.5% . In the cubic phase at higher pressures an orbitally liquid state is preferred. The two phases differ in their bulk modulii with the low-pressure phase being highly compressible compared to the high-pressure cubic phase, thereby explaining the large volume collapse. The calculated volume collapse is one of the largest known in transition metal oxides. Presence of such a ferroelastic transition with polar displacements in a magnetic system is interesting and could be exploited to make magnetostrictive materials.

We thank Dr. Ho-kwang Mao in GL for useful suggestions and Dr. Jian Xu and Dr. Wensheng Xiao for sharing their experimental findings with us. This work was partly supported by the EFree, an Energy Frontier Research Center funded by the U.S. Department of Energy, Office of Science, Office of Basic Energy Sciences under Award Number DE-SC0001057 and partly by the Office of Naval Research No. N00014-07-1-0451 . Part of the computation was performed at the Center for Piezoelectrics by Design, College of William and Mary.

**Figure 1.** Total energy difference with GGA+U (U=3eV and J=0.87 eV) as a function of Pb displacements in an otherwise cubic structure with a=4.0Å (i.e.$V_0$=64 Å$^3$) and

a=3.713Å i.e. 20% compression with respect to $V_0$ (in red).   Without compression Pb off-centering is energetically favorable with displacements along the tetragonal direction giving the lowest energy.   Under a 20% compression the cubic phase has the lowest energy.

**Figure 2.** Energy vs volume curve for the cubic and the fully relaxed tetragonal structure using GGA+U and LDA+U**.** (Inset **a**) Data from Figure 4 of Ref. [11] was inverted to obtain the energy vs volume curve for SrCrO$_3$. (Inset **b**) Enthalpy of the two phases using GGA+U showing a predicted transition pressure of ~ 7GPa and the accompanying volume change of ~ 11.5%.

**Figure 3.** Band structure plots for the tetragonal (*P4mm*) and the cubic (*Pm-3m*) phases at the experimentally measured anomalously large volume under ambient conditions of 64.0 Å$^3$.   Fat bands are plotted for the majority-spin Cr-d orbitals. The insets show corresponding partial density of states plots with units of states/eV/cell from a separate FP-LMTO calculation. There is large splitting of the $t_{xy}$ orbitals in the tetragonal phase compared to the cubic phase, and they form a dispersionless band. This leads to full occupancy of this orbital.   This orbital ordering is responsible for stabilizing the experimentally observed large volume in ambient conditions.

**T**able 1. Equation of state parameters from GGA+U calculations for the low pressure tetragonal phase and the high pressure cubic phase.   The equilibrium volume of the tetragonal phase is in good agreement with the experimentally measured large volume ~ 64 Å$^3$.   Theory further suggests a large difference in the bulk modulus between the two phases, giving rise to a large volume collapse across the transition.

| Structure | EOS parameters |
|---|---|
| (low pressure) Tetragonal phase (*P4mm*) | $V_0$ = 66.14 Å$^3$ |
|  | $K_0$ = 104 GPa |
|  | $K_0$' = 4 |
| (high pressure) Cubic phase (*Pm-3m*) | $V_0$ = 57.34 Å$^3$ |
|  | $K_0$ = 167 GPa |
|  | $K_0$' = 4 |

**Figure 1:**

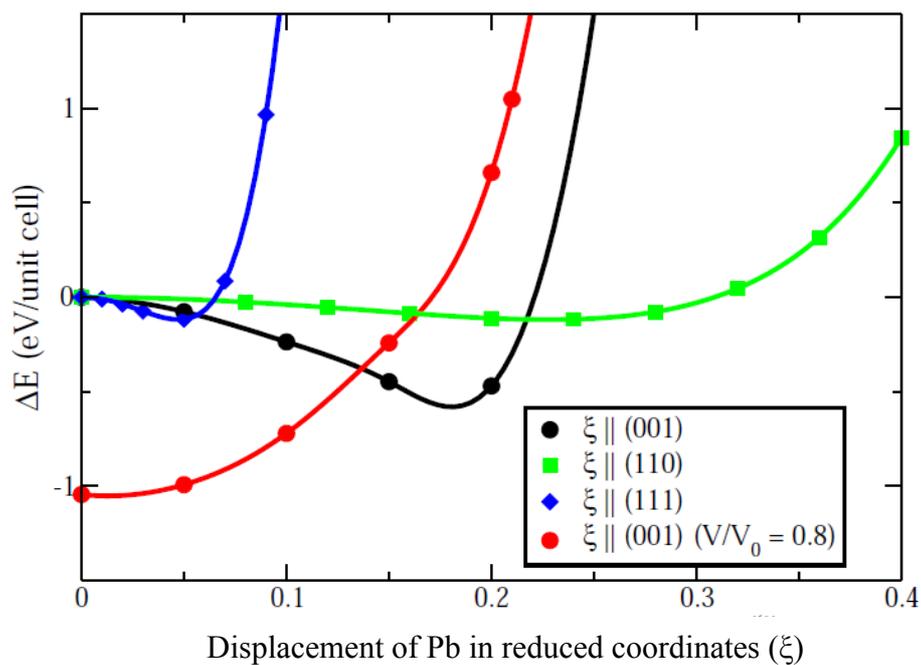

**Figure 2:**

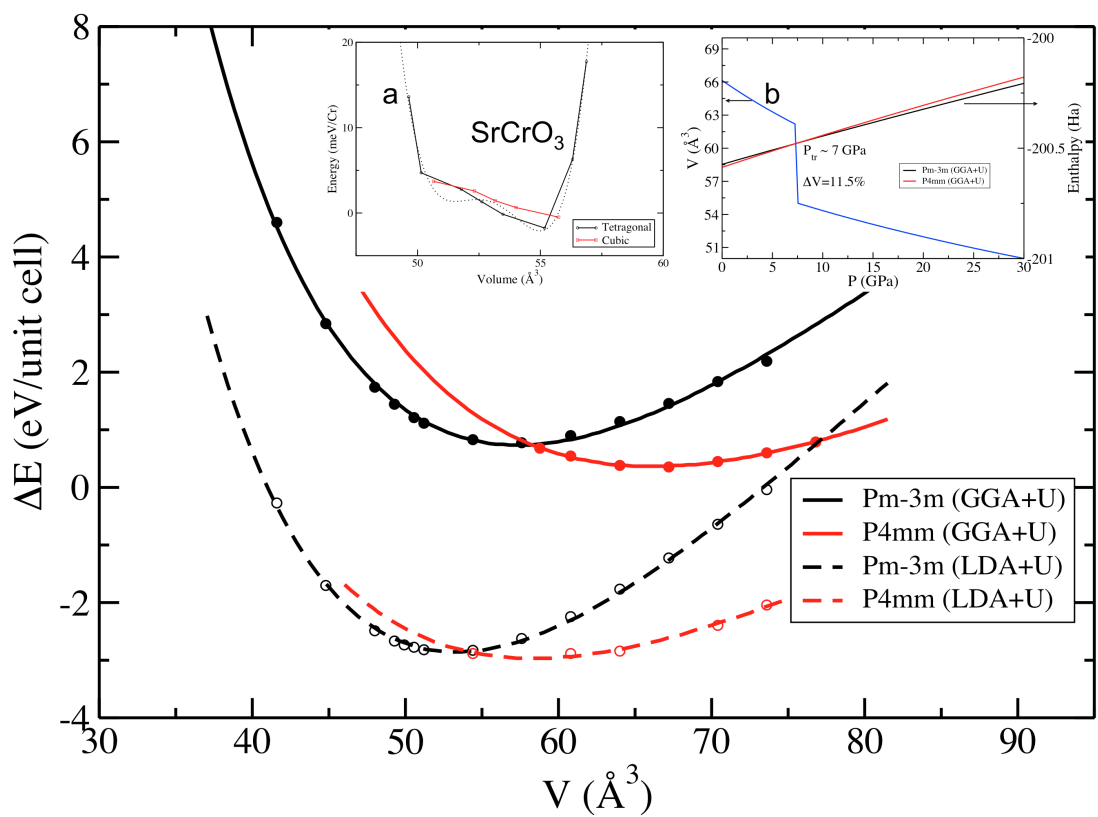

**Figure 3:**

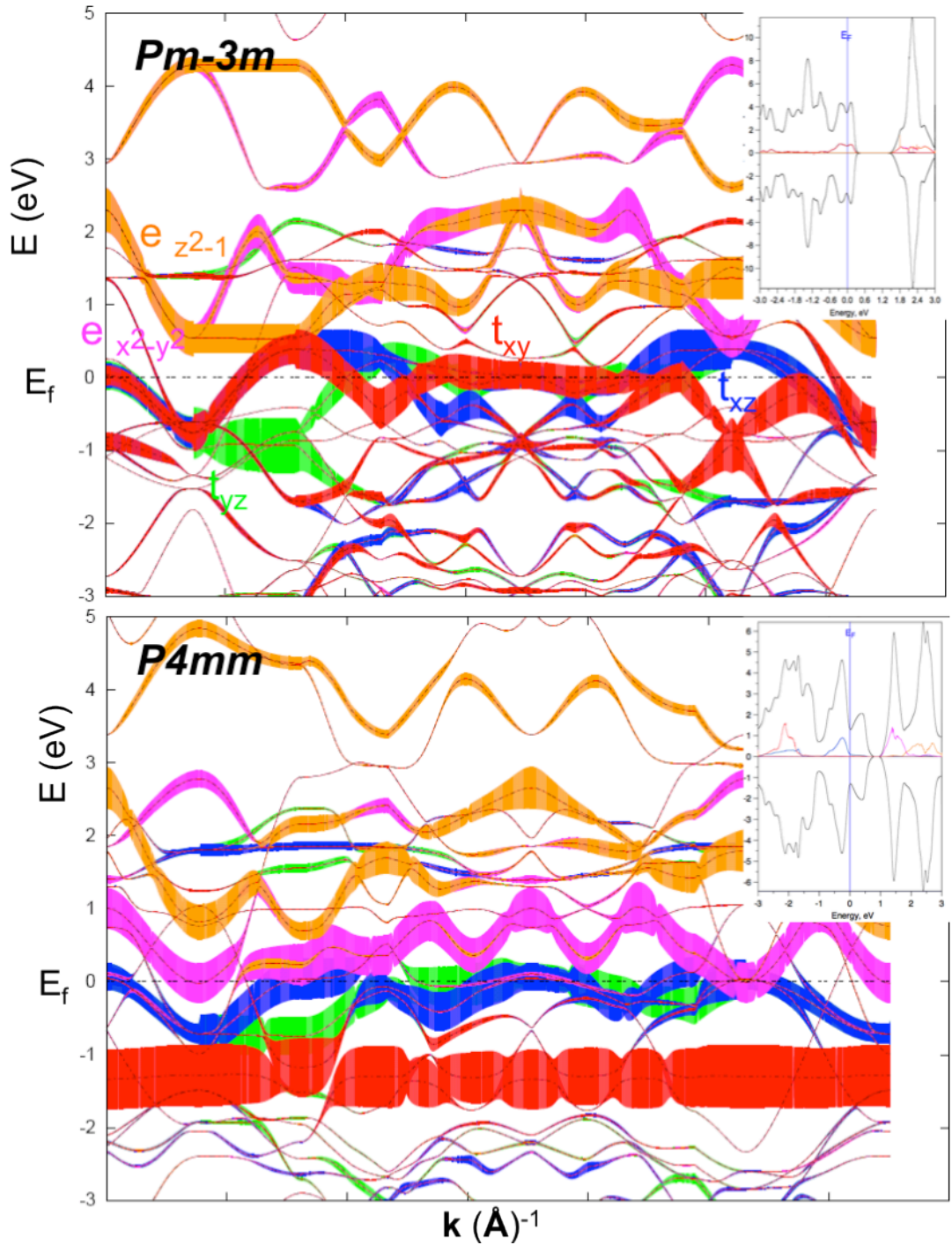